\documentclass[11pt]{article}
\usepackage{mathrsfs}
\usepackage{amsmath}
\usepackage{amssymb}
 \allowdisplaybreaks[4]
\makeatother       

\makeatletter      
\@addtoreset{equation}{section}
\makeatother       

\title{ \Huge  Algebraic constraints on Tau functions of P reduced KP with String equation}
\author{  Liu Shaowei\dag\thanks{Corresponding author, Email: swliu@mail.ustc.edu.cn},\\
\dag\scriptsize Department of  Mathematics, University of Science
and Technology of China, Hefei, 230026 Anhui,P.R. China \\
 }
 \vspace{1cm}
\date{}
\begin{document}
\maketitle
\begin{center}
\begin{center}{\bf Abstract}\end{center}
\vspace{1cm}
\begin{minipage}{12cm}
 { Give a convenient new approach to obtain the algebraic constraints on tau functions of
 p-reduced KP with the constraints of string equation. The classical results about algebraic constraints
 on associated tau function are included.}
 \end{minipage}
\end{center}

\vspace{2cm}
\newpage
\section*{\S 1.\ \ Introduction}
\setcounter{section}{1}

Ed.Witten and Kontsevich\cite{k,w1,w2} gave a classical result that
solutions of KdV equation with the constraint of string equation
corresponds to partition function of 2D quantum gravity or
generating function of intersection number. And Kontsevich\cite{k}
showed the associative solution by matrix integral. Since the
solution of the KdV equation can be characterized by a single
function of tau function\cite{jim}, it also showed that the tau
function is  a vacuum vector for the Virasoro Algebra. Since
integrable system has tightly connection with string theory and
intersection theory\cite{mo,di}, the results can be show in language
of integrable system\cite{dk4,dk3,am1}. Also it was considered to
extend the results from 2-reduced KP to p-reduced KP hierarchy, and
conjecture\cite{dvv,fkn} that for general p the tau function which
satisfy the string equation and p-reduced KP hierarchy are
equivalent to the vacuum vector of  a $W_p$-algebra. Goeree\cite{go}
shows that it is true when $p=3$. The case for higher p was also
researched\cite{fkn,mo,am1}. And extend the corresponding results to
other integrable hierarchy is also an interesting topic, such as for
BKP, qKP, mKP, etc.

In the paper we show a new convenient approach for obtain the
algebra constraints on tau functions of p-reduced KP constrained by
string equation. With this method we can naturally get the condition
of $t_{kp}=0$. And these algebraic constrains include the classic
results about Virasora constraints and $W_p^+$ constrains which is
the conjecture\cite{dvv}. The whole process of computation could
greatly simplify by the tool of the ASvM
formula\cite{dk3,asm2,asm3}.  In these process we found that we can
not directly let $t_{kp}=0$, which makes $t_{kp}=0$ as an additional
condition. And we also found for obtain strictly Virasoro, we should
add a constant on the item of constraint of $\mathcal {L}_0$, not
let all constant $c$=0. Fortunately we found the additional constant
just cancel with the item including $t_{kp}$ in the next higher
order algebraic constraint of $k=-1$. Inspire by that we extend
these method to the whole algebraic constraints and obtain our
results of theorem 3.2. And because the additional constant do not
destroy the algebra structure and we could stop the process at a
fixed number, we also obtain Theorem 3.3.

The organization of the paper is as follows. In section 2, for
self-contained we give a brief description of  KP hierarchy. In
section 3, we deduced the algebraic constraints on tau function.
\section*{\S 2.  KP hierarchy }
\setcounter{section}{2}

        To be self-contained, we list some properties of KP
hierarchy, based on detailed research in \cite{dk2}.
\[ f(t)=f(x+t_1,t_2,\cdots,t_j,\cdots),
\]
for any$j\in\mathbb{Z}$
\begin{equation}\partial^j\circ
f=\sum^{\infty}_{i=0}\binom{j}{i}(\partial^i
f)\partial^{j-i},\hspace{.3cm}
\binom{j}{i}=\frac{j(j-1)\cdots(j-i+1)}{i!}.\label{21}
\end{equation}
Define the KP operator as follow
\begin{equation}
L=\partial+ \sum_{j=1}^{\infty} f_j(t)\partial^{-j}.
\end{equation}
 Denote by $L_+=\sum_{j=0}^d f_j(n)\partial^j$, and by $L_-=\sum_{j=-\infty}^{-1}
f_j(n)\partial^j$.

         The KP-hierarchy \cite{dk2} is a family of evolution equation in
infinitely many variables $t=(x+t_1,t_2,\cdots)$
\begin{equation}
\frac{\partial L}{\partial t_i}=[(L^i)_+, L].\label{27}
\end{equation}

There are dressing operator $\Phi(t)$
\[\Phi(t)=1+\sum^\infty_{j=1}w_j(t)\partial^{-j},
\]
and
\begin{equation}
L=\Phi(t)\circ\triangle\circ \Phi(t)^{-1}.\label{22}
\end{equation}
         There are wave function $w(t,z)$ and adjoint wave
function $w^*(t,z)$,
\begin{align}
w(t,z)=\Phi(t)exp(\sum^\infty_{i=1}t_i z^i)
\end{align}
and
\begin{align}
w^*(t,z)&=(\Phi^{-1}(t))^* exp(\sum^\infty_{i=1}-t_i z^i).
\end{align}

 And a tau function $\tau(t)$  for KP
exists, which satisfies that
\begin{equation}
w(t,z)=\frac{\tau(t-[z^{-1}])}{\tau(t)} exp(\sum^\infty_{i=1}t_i
z^i)\label{24}
\end{equation}
and
\begin{equation}
w^*(t,z)=\frac{\tau(t+[z^{-1}])}{\tau(t)} exp(\sum^\infty_{i=1}-t_i
z^i),\label{25}
\end{equation}
where denote $[z]=(z,\frac{z^2}{2},\frac{z^3}{3},\cdots)$. Also
introduce the $G(z)$ operator, which action is $G_t(z)f(t)=f(t-[z])$
and $G_{t'}^*(z)f(t')=f(t'+[z])$.

There are vertex operators
\begin{equation}
X(\lambda,\mu)=:\exp\sum_{-\infty}^\infty(\frac{P_i}{i\lambda^i}-\frac{P_i}{i\mu^i}):
\end{equation}
where where
\begin{eqnarray}
  P_i=
  \begin{cases}
    \partial_i,\quad &i>0\\ \\
    |i|t_{|i|},\quad &i\leq 0
  \end{cases}\notag
\end{eqnarray}
And there are additional symmetries. Define
\begin{equation}
  \Gamma=\sum_{i=1}^\infty it_i\partial^{i-1},
\end{equation}
and
\begin{equation}
  M=\Phi(t)\circ\Gamma\circ \Phi(t)^{-1}.
\end{equation}
then the additional symmetries are defined by
\begin{equation}
  \partial^*_{ml}\Phi(t)=-(M^mL^l)_-\circ \Phi(t).
\end{equation}
Here for convenient the symbols have a slight different from
\cite{dk2}.

Talyor expand the $X(\lambda,\mu)$ in $\mu$ at the point of
$\lambda$, there are the following
\begin{equation}
X(\lambda,\mu)=\sum_{m=0}^\infty
\frac{(\mu-\lambda)^m}{m!}\sum_{n=-\infty}^\infty\lambda^{-n-m}W_n^{(m)}.
\end{equation}
Where
\[\sum_{n=-\infty}^\infty\lambda^{-n-m}W_n^{(m)}=\partial_\mu^m|_{\mu=\lambda} X(\lambda,\mu).
\]
and the first items are
\begin{align}
&W_n^{(0)}=\delta_{n,0},\notag\\
&W_n^{(1)}=P_n,\notag\\
&W_n^{(2)}=\sum_{i+j=n}:P_iP_j:-(n+1)P_n,\notag\\
&W_n^{(3)}=\sum_{i+j+k=n}:P_iP_jP_k:-\frac{3}{2}(n+2)\sum_{i+j=n}:P_iP_j:+(n+1)(n+2)P_n,\notag\\
&W_n^{(4)}=P_n^{(4)}-2(n+3)P_n^{(3)}+(2n^2+9n+11)P_n^{(2)}-(n+1)(n+2)(n+3)P_n\\
&\cdots\cdots
\end{align}
Here denote
\begin{align}
&P_n^{(2)}=\sum_{i+j=n}:P_iP_j:\\
&P_n^{(3)}=\sum_{i+j+k=n}:P_iP_jP_k:\\
&P_n^{(4)}=\sum_{i+j+k+l=n}:P_iP_jP_kP_l:-\sum_{i+j=n}:ijP_iP_j:\\
&\cdots\cdots
\end{align}
And there are ASvM formula for KP hierarchy\cite{dk3,asm2,asm3},
that is
\begin{equation}
\partial^*_{m,l+m}\tau(n;t)=\frac{W_l^{{(m+1)}}\cdot
\tau(t)}{m+1},
\end{equation}
which hold for $m\geq 0$ and for all $l$.
\section*{\S 3 Algebraic constraints on tau functions}
\setcounter{section}{3} \setcounter{equation}{0}

In this section we deduce the algebraic constraints on tau functions
which satisfy the p-reduced KP and sting equation.

First we should oobtain the action of additional symmetries on tau
functions from the action of additional symmetries on wave
functions. That is the following lemma

{\sl {\bf   lemma3.1}} For $m\geq 0$, and $\forall l$, there are
\begin{equation}
  \partial^*_{m,m+l}w(t,z)=(G(z)-1)
  \frac{W_l^{{(m+1)}}/m+1\cdot\tau(t)}{\tau(t)}\cdot w(t,z).
\end{equation}
{\sl {\bf   Proof:}} There are
\begin{align}
 &\partial^*_{m,m+l}w(t,z)\\=&\partial^*_{m,m+l}(\frac{G(z)\tau(t)}{\tau(t)}\exp\sum_{i=1}^\infty
 (t_iz^i))\notag\\
 =&\frac{\tau(t)\cdot G(z)(\partial^*_{m,m+l}\tau(t))-G(z)\tau(t)\cdot\partial^*_{m,m+l}\tau(t)}{\tau^2(t)}\exp\sum_{i=1}^\infty
 (t_iz^i)\notag\\
 =&((G(z)-1)\frac{W_l^{(m+1)}/m+1\cdot\tau(t)}{\tau(t)})w(t,z).\notag
\end{align}
$\hfill{\Box}$

Now consider the p-reduced KP and string equation. The string
equation \cite{dk2} means that find two differential operators
satisfy $[P,Q]=1$, that is
\begin{equation}
[L^p,\frac{1}{p}(ML^{-p+1})_+]=1.
\end{equation}which is equivalent the following two conditions. First
the p-reduced condition
\begin{equation}
(L^{kp})_-=0,
\end{equation}
which means $L$ is independent on $t_{kp}$. And that also means
$\tau(t)$ is independent on $t_{kp}$.

Second satisfy the following identities
\begin{equation}
  \partial^*_{1,-p+1}L=0.\label{61}
\end{equation}
We found that the condition of (\ref{61}) could lead to the
different values of the dressing operator $\Phi(t)$, that is these
different values of dressing operator can all keep the condition of
(\ref{61}). And the different values lead to different algebraic
constraints on tau functions. For example, if take
\begin{equation}
 \partial^*_{1,-p+1}\Phi(t)=0,
\end{equation}
it can satisfy the condition (\ref{61}) and obtain
\begin{equation}
\partial^*_{1,-p+1}w(t,z)=0.
\end{equation}
And from the lemma 3.1, it obtain
\begin{equation}
  \frac{W_{-p}^{(2)}\cdot\tau(t)}{2}=c\tau(t).
\end{equation}
that is
\begin{equation}
(\frac{1}{2}\sum_{i+j=-p}:P_iP_j:-(-p+1)P_{-p})\cdot\tau(t)=c\tau(t).
\end{equation}
but because the left hand $\frac{W_{-p}^{(2)}\cdot\tau(t)}{2}$
include the variable $t_p$, but the right hand $\tau(t)$ does not it
lead to that no solution of the $\tau(t)$ functions satisfy the
above identity.

So to keep the existence of the solution ,we let
\begin{equation}
 \partial^*_{1,-p+1}\Phi(t)=-\frac{p-1}{2}L^{-p}\label{63},
\end{equation}
it also keep the equation (\ref{61}), but it obtain
\begin{equation}
\partial^*_{1,-p+1}w(t,z)=-\frac{p-1}{2}z^{-p}\cdot w(t,z).
\end{equation}
Use lemma 3.1, that is
\begin{equation}
  (G(z)-1)\frac{W_{-p}^{(2)}/2\cdot\tau(t)}{\tau(t)}=-\frac{p-1}{2}z^{-p}.
\end{equation}
which means
\begin{equation}
  (G(z)-1)\frac{\mathcal {L}_{-p}\cdot\tau(t)}{\tau(t)}=0.
\end{equation}
That is
\begin{equation}
\frac{1}{2}\sum_{i+j=-p}:P_i P_j:\tau(t)=\mathcal
{L}_{-p}\tau(t)=c\tau(t),\label{65}
\end{equation}
then obtain the same results in \cite{dk4}.

Note that the identity of (\ref{63}) is very important, and we can
obtain the whole algebraic constraints only from it and the
p-reduced condition. So (\ref{65}) is also the basic equation that
tau function should satisfy.

With the definition of additional symmetries the condition
(\ref{63}) is equivalent to
\begin{equation}
  (ML^{-p+1})_-=\frac{p-1}{2}L^{-p}.\label{64}
\end{equation}
And only start with the (\ref{64}), together with the p-reduced
conditions, there are a general results\cite{am1}
\begin{align}
(M^jL^{kp+j})_-&=\prod_{r=0}^{j-1}(\frac{p-1}{2}-r)L^{-p}\quad
&k=-1\quad j=1,2,\cdots\\
&=0\quad &k=0,1,2,\cdots \quad j=1,2,\cdots
\end{align}
In term of dressing operators that is
\begin{align}
\partial^*_{j,kp+j}\Phi(t)&=-\prod_{r=0}^{j-1}(\frac{p-1}{2}-r)L^{-p}\circ\Phi(t)\quad
&k=-1\quad j=1,2,\cdots\\
&=0\quad &k=0,1,2,\cdots \quad j=1,2,\cdots
\end{align}
And In term of the wave functions that is
\begin{align}
\partial^*_{j,kp+j}w(t,z)&=-\prod_{r=0}^{j-1}(\frac{p-1}{2}-r)z^{-p}\quad
&k=-1\quad j=1,2,\cdots\\
&=0\quad &k=0,1,2,\cdots \quad j=1,2,\cdots\label{68}
\end{align}
And with the above methods of ASvM, there are
\begin{align}
(G(z)-1)\frac{W_{kp}^{{(j+1)}}/j+1\cdot\tau(t)}{\tau(t)}&=-\prod_{r=0}^{j-1}(\frac{p-1}{2}-r)z^{-p}\quad
&k=-1\quad j=1,2,\cdots\notag\\
&=0\quad &k=0,1,2,\cdots \quad j=1,2,\cdots\label{69}
\end{align}
These are the equations which the tau functions under the
constraints of p-reduced conditions and string equation should
satisfy. And these equations are equivalent to (\ref{63}) and the
reduce conditions. And we could deduce the algebraic constraints on
tau function only from the above equations. Notice that the item of
$k=-1$ play an important role in above equation.

Notice that the above equation lead to a lot of uncertain constant,
and the differen value of these constants may lead to different
algebraic structure. Here we show a method to obtain a algebraic
structure including the classical results.

Start with $k=-1$, $j=1$, that is the result (\ref{65}). In all we
obtain
\begin{align}
&\mathcal {L}_{-p}\cdot \tau(t)=c\tau(t)\quad &k=-1\quad j=1\\
&\frac{W_{kp}^{(2)}}{2}\cdot \tau(t)=c \tau(t)\quad &k=0,1,2,\cdots
\quad j=1\label{66}
\end{align}
And noticed that when $k\geq 0$, the $P_{kp}=\partial_{kp}$ or
$P_{kp}=0$. And because the p-reduced conditions the $\tau(t)$ is
independent on $t_{kp}$, there are
\begin{equation}
P_{kp}\cdot\tau(t)=0\quad k=0,1,\cdots
\end{equation}
So in (\ref{66}) we can get rid of the items of $P_{kp}$ in
$W^{(2)}_{kp}$. Then the remain of (\ref{66}) are $\mathcal
{L}_{kp}$. So in all it obtains
\begin{equation}
  \mathcal {L}_{kp}\cdot\tau(t)=c\tau(t)\quad k=-1,0,1,2,\cdots
\end{equation}
Here c is a constant. With the expression of $\mathcal {L}$ notice
that under the condition of $\partial_{kp}\tau(t)=0$ the $\mathcal
{L}_{kp}$ can naturally get rid of the items which include the
variable $t_{kp}$. So we need not to let $t_{kp}=0$ in special or as
some additional conditions. It was satisfied naturally. For
convenient also denote $\mathcal {L}|_{t_{kp}=0}$ by $\mathcal {L}$.

If we want to get the Virasoro constraints, we can not simply let
all $c=0$ because the $\mathcal {L}_{kp}$ do not form the virasoro.
The Virasoro satisfy the commutatiom
\begin{equation}
[\mathcal {L}_n, \mathcal {L}_m]=(n-m)\mathcal
{L}_{n+m}+\frac{1}{12}(n^3-n)\delta_{n+m},
\end{equation}
but the $\mathcal {L}_{kp}$ satisfy
\begin{equation}
[\mathcal {L}_{np}, \mathcal {L}_{mp}]=(np-mp)\mathcal
{L}_{np+mp}+\frac{1}{12}((np)^3-np)\delta_{np+mp}.
\end{equation}
So to get the Virasoro constrains we should firstly let
\begin{equation}
  L_n=\frac{1}{p}\mathcal {L}_{np}\quad n=-1,0,1,2,\cdots
\end{equation}
but it does not enough. Secondly to keep the action of $\mathcal
{L}_0$ on tau function is correct, that is to keep
\begin{equation}
[\mathcal {L}_{-p},\mathcal {L}_p]\tau(t)=0,
\end{equation}we should let
\begin{equation}
  \mathcal {L}_0\tau(t)=-\frac{1}{24}(p^2-1)\tau(t),\label{67}
\end{equation}
which means for $\mathcal {L}_0$ we take constant
$c=-\frac{1}{24}(p^2-1)$ and for other $k=-1,1,2,\cdots$, let $c=0$
which is
\begin{equation}
  \mathcal {L}_{kp}\tau(t)=0\quad k=-1,1,2,\cdots
\end{equation}
The change of (\ref{67}) is very important in higher order algebraic
constraints and we shall show it later. Then take
\begin{equation}
  L_0=\frac{1}{p}\mathcal {L}_0+\frac{1}{24p}(p^2-1),
\end{equation}
and because the constant do not effect the Poisson Bracket there are
correct Virasoro relation
\begin{equation}
[L_n, L_m]=(n-m)L_{n+m}+\frac{1}{12}(n^3-n)\delta_{n+m},
\end{equation}
and
\begin{equation}
L_n\tau=0 \quad n=-1,0,1,\cdots
\end{equation}
So we get the Virasoro constraints as a subalgebra. For $p=2$, that
is the classical results\cite{k,w1,w2}.

Now go on this process. consider $j=2$ in equations (\ref{69}) which
lead to
\begin{align}
&(G(z)-1)\frac{\frac{1}{3}(P_{kp}^{(3)}-\frac{3}{2}(kp+2)P_{kp}^{(2)}+(kp+1)(kp+2)P_{kp})\cdot\tau(t)}{\tau(t)}\notag\\
&=-\frac{p-3}{2}\cdot\frac{p-1}{2}z^{-p}\quad
k=-1\quad j=1,2,\cdots\notag\\
&=0\quad\quad\quad\quad k=0,1,2,\cdots \quad j=1,2,\cdots\label{610}
\end{align}
First consider $k=-1$, with the results of $j=1$, there are
\begin{align}
&(G(z)-1)\frac{\frac{1}{3}(P_{-p}^{(3)}+(p-1)(p-2)P_{-p})\cdot\tau(t)}{\tau(t)}=-\frac{p-3}{2}\cdot\frac{p-1}{2}z^{-p}\label{611}
\end{align}
Also same to $j=1$, to keep the existence of the solution of tau
function, there should get rid of the items which include the
variables $t_{kp}$. But we can not let $t_{kp}=0$ directly which is
as a additional conditions. we should get rid of them naturally.
Notice that in $P_{-p}^{(3)}$, the items include $t_{kp}$ is $3\cdot
P_{-kp}\cdot\sum_{i+j=kp-p}P_iP_j$. Use the results of $j=1$, we
only need consider $3\cdot P_{-p}\cdot\sum_{i+j=0}P_iP_j$. So
consider the the change of (\ref{67}), the equation (\ref{611}) jus
show that
\begin{equation}
  (P_{-p}^{(3)}|_{\text{without}\quad t_{kp})}\cdot \tau(t)=c\tau(t),
\end{equation}
which means the the change of (\ref{67}) just cancels with the items
include $t_p$. So let $c=0$ there are
\begin{equation}
  P^{(3)}_{-p}|_{t_{kp}=0}\cdot \tau(t)=0.
\end{equation}
As for the equations of $k>0$, we can also naturally get rid of the
items which include the $t_{kp}$ by the results of $j=1$. Also let
the corresponding $c=0$, there are
\begin{equation}
  P^{(3)}_{kp}|_{t_{kp}=0}\cdot \tau(t)=0\quad k=1,2,\cdots
\end{equation}
As for the item of $k=0$, the equation is
\begin{equation}
(G(z)-1)\frac{\frac{1}{3}(P_0^{(3)}-3P_0^{(2)})\cdot\tau(t)}{\tau(t)}=0,
\end{equation}
for satisfy the equation and by the same analysis, there is
\begin{equation}
  P_0^{(3)}|_{t_{kp}=0}\cdot\tau(t)=c\tau(t).
\end{equation}
Now we should fix on the constant $c$. If we want to obtain the next
higher order algebraic constraints, we do not give $c$ an arbitrary
value. If we stop the process here, just let $c=0$. Otherwise notice
that the change of (\ref{67}) and inspire by that. We found that if
we give $c$ a appropriate value the higher the equation $j=3,k=-1$
could holds, which also means that we can obtain the value of $c$
from the next higher algebraic constraint equation of $k=-1$. Here
for $j=2,k=0$, $c=0$, we shall show it in the following.

Continue the process, consider the case of $j=3$,
\begin{align}
&(G(z)-1)\frac{1}{\tau(t)}\cdot\notag\\
&\frac{1}{4}(P_{kp}^{(4)}-2(kp+3)P_{kp}^{(3)}+(2(kp)^2+9(kp)+11)P_{kp}^{(2)}-(kp+1)(kp+2)(kp+3)P_{kp})\cdot\tau(t)\notag\\
&=-\frac{p-5}{2}\cdot\frac{p-3}{2}\cdot\frac{p-1}{2}z^{-p}\quad
k=-1\quad j=1,2,\cdots\notag\\
&=0\quad\quad\quad\quad k=0,1,2,\cdots \quad j=1,2,\cdots\label{610}
\end{align}
we can also use the method above and the same analysis and the
results of $j=1,2$ to obtain
\begin{equation}
  P_{kp}^{(4)}|_{t_{kp=0}}\cdot\tau(t)=c\tau(t)\quad k=-1,0,1,2,\cdots
\end{equation}
And here the items include the variables $t_{kp}$ also can naturally
be get rid of.

And in the process of obtain the result of
\begin{equation}
P_{-p}^{(4)}|_{t_{kp}=0}\cdot\tau(t)=c\tau(t),
\end{equation}
which comes from the equation
\begin{align}
&(G(z)-1)\frac{1}{\tau(t)}\cdot
\frac{1}{4}(P_{-p}^{(4)}+2(p-3)P_{-p}^{(3)}+(p-1)(p-2)(p-3)P_{kp})\cdot\tau(t)\notag\\
&=-\frac{p-5}{2}\cdot\frac{p-3}{2}\cdot\frac{p-1}{2}z^{-p}.\label{612}
\end{align}
Notice that from the change of (\ref{67}),there is
\begin{equation}
\sum_{i+j=0}P_iP_j=-\frac{1}{12}(p^2-1).
\end{equation}

For get rid of the items include of $t_p$, there should have the
following simple equation
\begin{equation}
c-\frac{(p^2-1)(p-3)}{8}+\frac{(p-1)(p-3)(p-3)}{4}=\frac{(p-1)(p-3)(p-5)}{8}\label{62},
\end{equation}
here $c$ is the constant $c$  for
\begin{equation}
P_0^{(3)}|_{t_{kp}=0}\cdot\tau(t)=c\tau(t).
\end{equation}
That means for satisfying the next higher order algebraic constrains
we can obtain the $c$ of $j=2,k=0$ by a simple equation (\ref{62}).
That is the method how to obtain the value of $c$ of $k=0$ for all
$j$. The solution of (\ref{62}) is $0$, which is the result we
mentioned above.

 Also we could let $c=0$, for $k=-1,1,,2\cdots$. And
the $c$ for item of $j=3,k=0$ could also be obtained from the next
higher order algebraic constraints equation by the same method. That
is to solve a simple equation similar to (\ref{62}). Also we can
stop here and just let the $c=0$. So we obtain the algebraic
constrain of $j=3$.

And we can continue the process for $j=4,5,\cdots$ with similar
simple analysis and computation.

Finally there are

{\sl {\bf   Theorem 3.2}} If $\tau((t)$ is vacuum vector of the
following algebra
\begin{align}
&P_{kp}|t_{kp}=0\cdot\tau(t)=0 \quad k=0,1,2,\cdots \\
&P^{(j)}_{kp}|t_{kp}=0\cdot\tau(t)=0\quad k=-1,1,2,\cdots \quad j=2,3,\cdots\\
&(P^{(j)}_0+c_j(p))|t_{kp}=0\cdot\tau(t)=0\quad j=2,3,\cdots\quad
c_2(p)=\frac{1}{12}(p^2-1),c_3(p)=0,\cdots
\end{align}
then $\tau(t)$ is the tau function of p-reduced KP and string
equation.

And as $j$ goes bigger the constrains goes more. Because the whole
algebra constraints just base on the two basic conditions, p-reduced
and (\ref{63}). So to make sure the existence of the solution, we
could stop this process at a fixed number, for example at $j=p$. and
the existence of the solutions already proved by Kontsevich\cite{k}
for $p=2$, and Adler\cite{am1} for general p. And notice that the
constant do not destroy the Poisson Bracket and the algebraic
structure. So there are

{\sl {\bf Theorem 3.3}} If $\tau((t)$ is vacuum vector of the
following $W_p^+$ algebra
\begin{align}
&P_{kp}|t_{kp}=0\cdot\tau(t)=0 \quad k=0,1,2,\cdots \\
&P^{(j)}_{kp}|t_{kp}=0\cdot\tau(t)=0\quad k=-1,1,2,\cdots \quad j=2,3,\cdots,p\\
&(P^{(j)}_0+c_j(p))|t_{kp}=0\cdot\tau(t)=0\quad j=2,3,\cdots,p\quad
c_2(p)=\frac{1}{12}(p^2-1),c_3(p)=0,\cdots
\end{align}
then $\tau(t)$ is the tau function of p-reduced KP and string
equation. That is the result of the conjecture\cite{dvv}.



\end{document}